# The first cryogenic DT layered, beryllium capsule implosion at the National Ignition Facility

D. C. Wilson[1], J. L. Kline[1], S. A. Yi[1], A. N. Simakov,[1] G. A. Kyrala[1], R. E. Olson,[1] T. S. Perry[1], F. E. Merrill[1], S. Batha[1], A.B.Zylstra[1],D. A. Callahan[2], W. Cassata[2], E. L. Dewald[2], S. W. Haan[2], D. E. Hinkel[2], O. A. Hurricane[2], N. Izumi[2], T. Ma[2], A. G. MacPhee[2], J. L. Milovich[2], J. E. Ralph[2], J. R. Rygg[2], M. B. Schneider[2], S. Sepke[2], D. J. Strozzi[2], R. Tommasini[2], C. Yeamans[2], H. G. Rinderknecht[3], B. H. Sio[3], MIT.

[1]Los Alamos National Laboratory, [2]Lawrence Livermore National Laboratory, [3]Massachusetts Institute of Technology

*National Ignition Facility experiments with beryllium capsules have followed a path begun with the highly successful "high-foot" plastic capsule implosion. Three shock timing keyhole targets, one low energy symmetry capsule, one backlighted symmetry capsule (1DConA viewed through a slit), and a 2DConA had been fielded leading up to a DT layered shot. In addition to backscatter subtraction, laser drive multipliers were needed to match observed X-ray drives. Those for the picket (0.95), trough (1.0) and second pulse (0.80) were determined by VISAR measurements. The time dependence of the Dante total x-ray flux and its fraction > 1.8 keV reflect the time dependence of the multipliers. A two step drive multiplier for the main pulse can match implosion times, but Dante measurements suggest the drive multiplier must increase late in time. With a single set of time dependent, multi-level multipliers the Dante data for shots N150222, N150410, and N150617 are well matched. Furthermore these same 3rd pulse drive multipliers also match the implosion times and Dante signals for plastic capsule DT shots N130501 and N130812. One discrepancy in the calculations is the X-ray flux in the picket, i.e. the first 3ns of the laser pulse. Calculations over-estimate the flux > 1.8 keV by a factor of ~100, while getting the total flux correct. These harder X-rays cause an expansion of the beryllium/fuel interface of 2-3 km/s before the arrival of the first shock. VISAR measurements show only 0.2 to 0.3 km/s. The X-ray drive on the DT capsule (N150617) was further degraded by a random decrease of 9% in the total picket flux. This small change in the total laser pulse energy caused the capsule fuel to change from an adiabat of 1.8 to 2.3 by mis-timing of the first and second shocks.*

*With this shock tuning and adjustments to the calculation, the first NIF beryllium capsule implosion achieved 29% of calculated yield, comparable to plastic DT capsules of 68% (N130501) and 21% (N130812). Inclusion of a large M1 asymmetry in the DT ice layer and*

*mixing from instability growth may help explain this final degradation. In summary when driven similarly the Be capsules performed like CH capsules. Performance degradation for both seems to be dominated by drive and capsule asymmetries.*

## Introduction

Beryllium ablators [1] covered the capsule of choice for the first ignition target design for the National Ignition Facility [2]. But the challenges of handling beryllium contamination in the target chamber and diagnostics led to the choice of glow discharge polymer (GDP), i.e. plastic, as the first capsule material used for ignition targets. Since 2009 hundreds of DT filled plastic capsules have been imploded at NIF, reaching neutron yields of nearly 1e+16, and demonstrating the effects of alpha particle heating [3], but not yet achieving ignition. The advantages of beryllium's lower opacity, higher ablation pressure, and higher density than plastic are well known [4,5,6], but only in 2014 were procedures in place and tested for monitoring beryllium. The construction of beryllium capsules was first accomplished by machining and bonding two hemi-shells [7]. More recently LLNL and General Atomics developed sputter deposition on plastic mandrels [8] with the mandrel then heated and removed as a gas. This is the technique that was used in the NIF implosions described here.

The first implosion of a DT filled beryllium capsule at NIF required a sequence of shots to understand the interaction of the beryllium capsule with the hohlraum and to experimentally tune the timing and velocity of shocks within the capsule. The first beryllium DT capsule experiment was designed to be similar to a plastic capsule tested on May 1, 2013 (N130501) [9]. Figure 1 shows the radii and composition of both capsules. Figure 2 a shows the two laser pulses, and figure 2b the radiation drive temperature histories. Both capsules were fielded in what is now a standard, 5750 µm diameter Au hohlraum. The laser entrance hole is slightly larger (3461 for beryllium vs 3101 µm dia. for CH) which is calculated to improve the drive symmetry. Both capsules were supported by a 45 nm thick tent and filled thru a similar metal and glass tube attached to a 10 µm dia. hole in the capsule. This similarity in laser drive, hohlraum dimensions, and capsule parameters allows us to compare the performance of these two experiments, and the performance of beryllium vs plastic capsules in general.

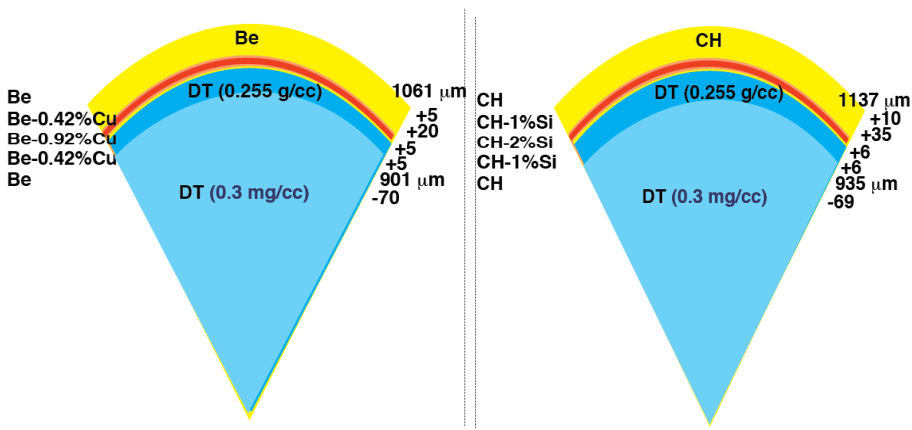

**Figure 1. Capsule diagrams for the Be (N150617), and plastic (N130501) capsules.**

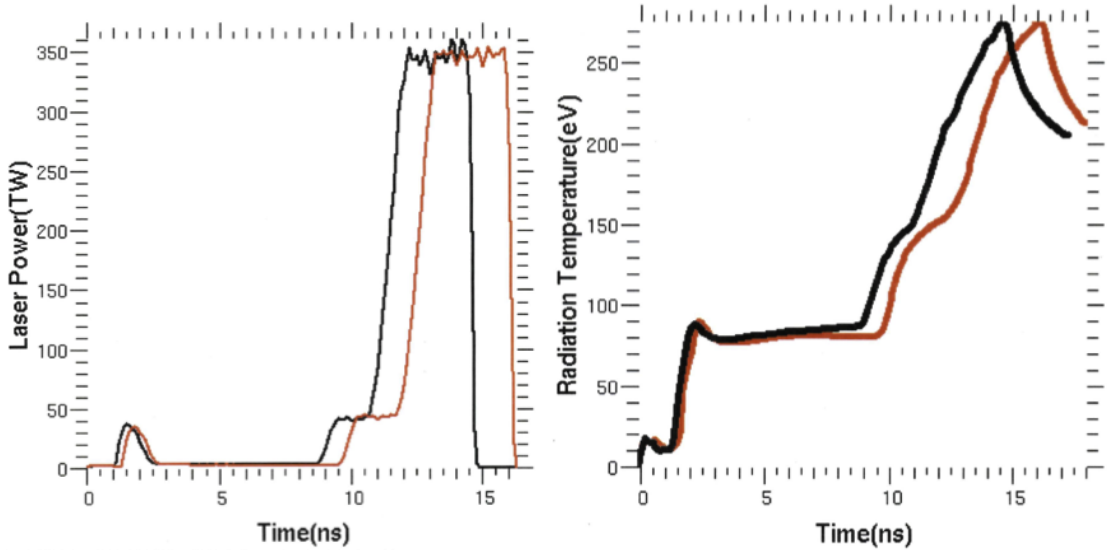

**Figure 2. Laser pulses and calculated hohlraum radiation temperatures for the Be (N150617, red), and plastic (N130501, black) DT shots.**

# X-ray Drive

Our HYDRA [10] calculations of beryllium capsule implosions have followed the outline described by Jones *et al*. [11]. The first step is to treat the backscattered laser energy as never having reached the hohlraum, subtracting it off the incident laser. Table 1 compares the measured time integrated backscatter from two beryllium experiments (N150222 and N150420) with that from a hohlraum with a plastic capsule. The plastic capsules have ~ 2% less backscatter, mostly due to less SRS in the 30° beams and no SBS in the 50° beams. Figures 3a,b, and c show the time history of this backscatter in hohlraums with Be (N150617) or CH (N130501) capsules with DT fill.

**Table I- Comparison of Hohlraum Backscatter with Beryllium and Plastic Capsules**

|  | Be | Be | CH |
|---|---|---|---|
| **Shot** | N150222 | N150420 | N130409 |
| **Laser Energy (MJ)** | 1.329 | 1.347 | 1.261 |
| **23° SRS** | 27% | 30% | 28% |
| SBS | 0.4% | 0.4% | 0.5% |
| **30° SRS** | 38% | 41% | 36% |
| SBS | 0.4% | 0.4% | 0.5% |
| **44° SRS** | 2.2% | 0.8% | 0.4% |
| SBS | 0.6% | 0.7% | 0.2% |
| **50° SRS** | 1.9% | 0.7% | 0.3% |
| SBS | 3.1% | 2.9% | 0.2% |
| **Total Backscatter** | 13.5 ± 3.2% | 13.7 ±1.8 | 11.4% |

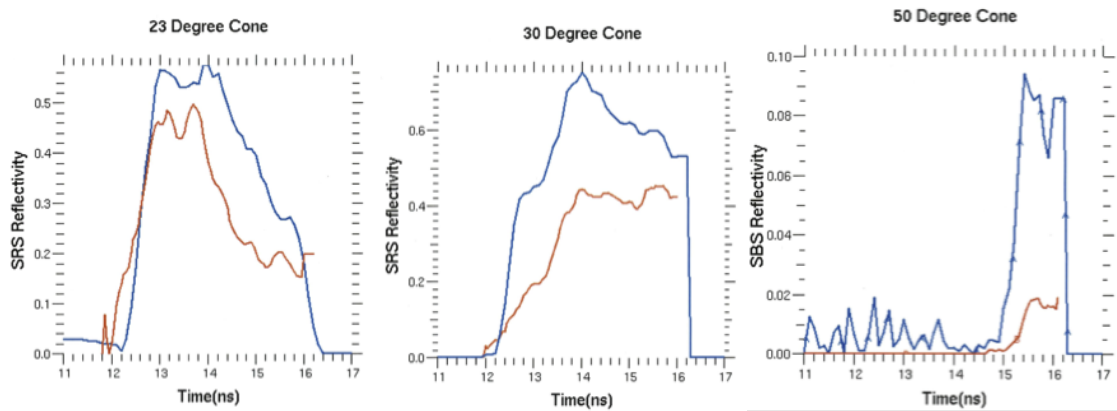

**Fig. 3. Time dependent SRS and SBS from hohlraums with Be (N150617, blue) and CH (N130501).**

Keyhole targets have a gold cone penetrating the hohlraum wall and into a capsule filled with liquid $D_2$. A VISAR (Velocity Interferometer System for Any Reflector) reflected off the moving shock front measures its velocity versus time [12]. A small mirror allows the VISAR to measure both the capsule equator and pole motions. With the laser backscatter removed, calculated velocities and timings can be compared with the data. Figure 4 shows the measured shock velocities and first cut simulated values. Ad hoc laser drive multipliers and energy transfer between the laser cones are needed to bring the calculations into agreement with experiment. For example, the calculated velocity of the first shock is adjusted by a drive multiplier during the picket of the laser pulse to bring the calculations into agreement with the data (labeled 1); the combined first and second shock velocity by a drive multiplier in the second laser pulse (4); and the third pulse multiplier to force agreement with the velocity of all three shocks combined (5). Slight corrections to the relative laser power in the inner (23 and 30°) and outer (44 and 50°) beams are still needed to match pole and equator timing differences (shown in 2 and 3).

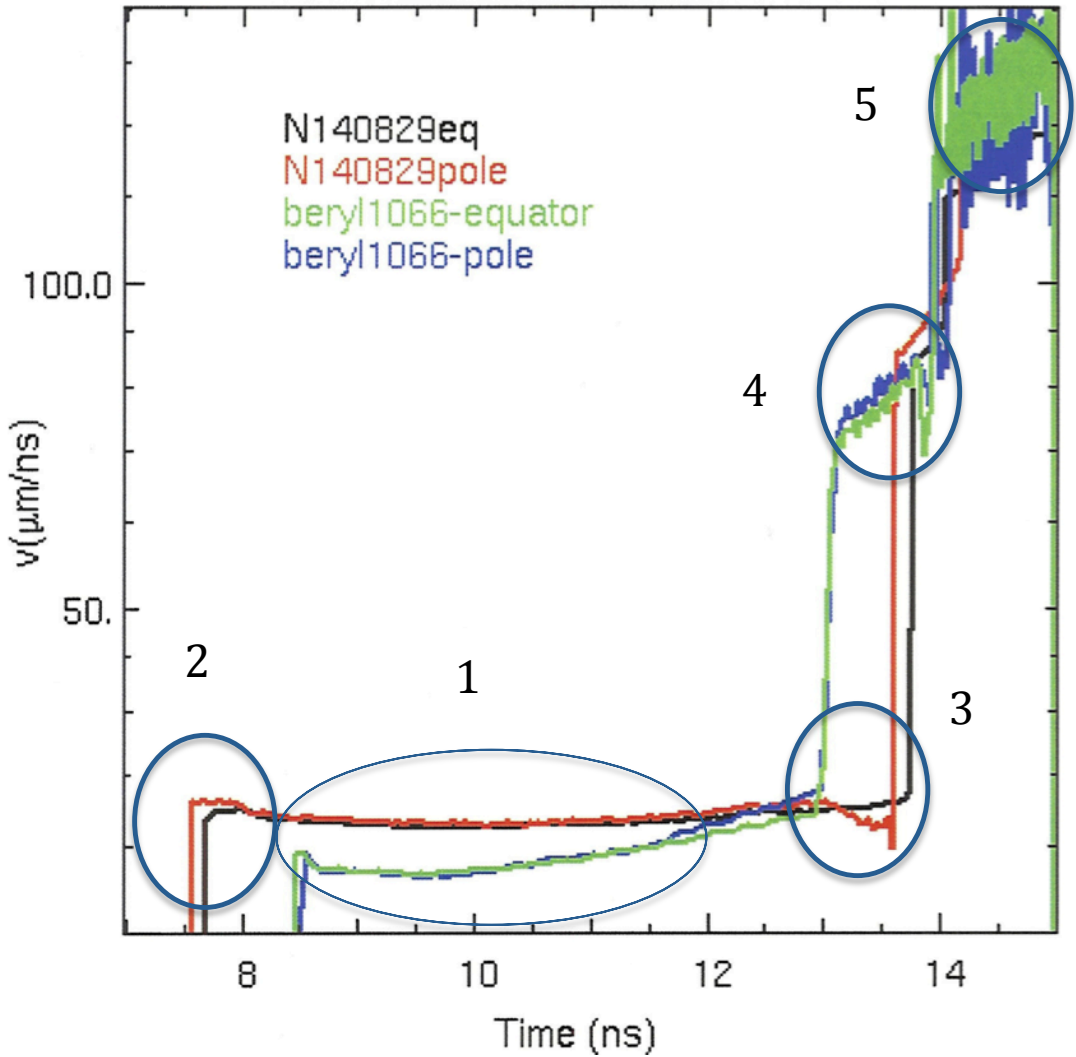

**Figure 4. Measured (red and black) and calculated VISAR shock velocities for N140829.**

In addition to the VISAR data, implosions backlighted with x-rays from a copper target gave streaked radius vs time (1DconA) [13] or time gated images of transmission (2DconA)[14]. Measurements of the time of peak X-ray emission at the south pole [15] (bangtime) added to the data constraining time dependent drive multipliers.

In post-shot modeling of NIF capsule implosions Jones *et al.* [11] introduced the application of time dependent multipliers on the laser power incident into the hohlraum as a calculational tool to match better the measured shock speeds, shock merger times, peak implosion velocity, capsule bangtime, and other experimental data. Experiments have shown that the radiation flux within the hohlraum is actually depleted as suggested by these laser drive multipliers [16]. The hohlraum physics in the calculations [17] is thought to be in error.

Figure 5 shows two sets of drive multipliers that are both consistent with the bangtimes, early VISAR, and backlit imaging results. Both use a 0.93 multiplier in the laser picket between about 1 and 3ns, and a 0.85 multiplier on the second pulse between about 9.5 and 11.5ns. These are required to match the 1st and 2nd shock timing and velocities from the VISAR data. The dip to 0.3 at 12ns is needed to explain the VISAR low 3rd shock velocity and the late bangtime in the low energy drive, 0.9 MJ, symmetry capsule implosion. The rise to either a constant value of 0.86, or an even higher rising values are required to match bangtimes from the 1DConA (N150222) and the 2DConA (N150420). The rise at the end of the pulse in the Multi-Step drive multipliers (red) is required to match Dante flux measurements as we will see later. The constant value in the Two Step curve (blue) is typical of modeling to match CH capsule performance.

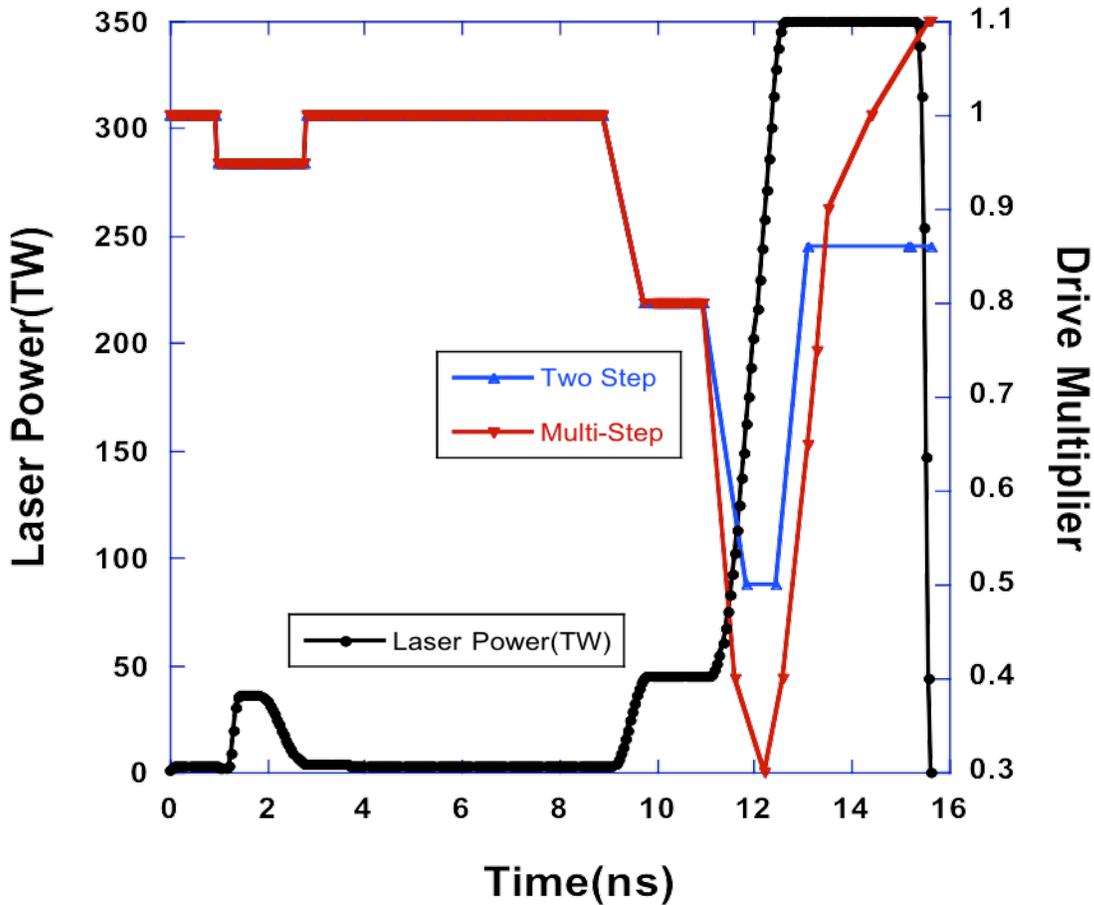

**Figure 5. Two sets of drive multipliers (red and blue) used in modeling the DT filled beryllium implosion (N150617) with its laser pulse.**

N150222 was a 1DconA target, using ~8.3 keV emission from a copper backlighter foil to measure in transmission the radius vs time of the imploding Be/Cu shell. From these the imploding velocity of the unablated mass is derived [13]. Figure 6 compares the experimental results (red dots) and calculations (blue dots) using the Multi-Step drive multiplier profile. The calculated points split into two curves because the code has difficulty defining the ablation front, and hence the mass remaining unablated. However there is good agreement within the error bars of calculated and observed velocities.

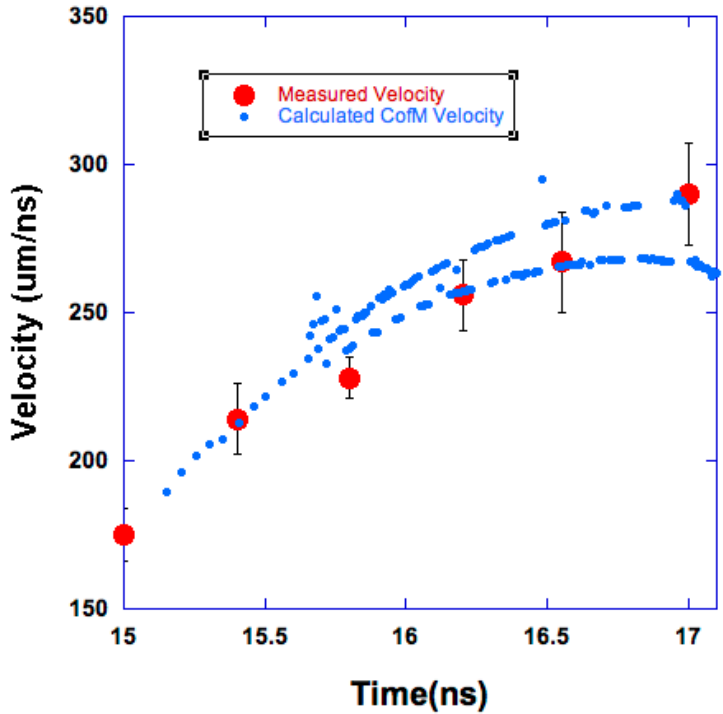

**Figure 6. Comparison between calculated (blue) and measured (red) center of mass implosion velocities for shot N150222.**

N150420 was a 2DconA target, again using ~8.3 keV copper radiation to obtain images of the implosion within about 1ns of peak X-ray emission (bangtime) [14]. From these time-dependent images radii of minimum transmission, and of the point of maximum slope in the transmission are derived. The former corresponds roughly to the radius on the inner beryllium layer doped with copper, and the latter to its outer radius. Figure 7 compares calculated radii with the measured. The agreement is quite good, but there appears to be a slight shift of the measured points to earlier times by about 30ps, within the 80ps uncertainty in the measured bangtime. The velocity derived from the measured minimum radii is 285 ± 20 um/ns compared with 324 um/ns calculated 0.62ns before bangtime. As with the 1DconA, N150222, the observed implosion is slightly early and slower than calculated. A calculated velocity for the maximum slope point is 245, also faster than the measured 223 km/s.

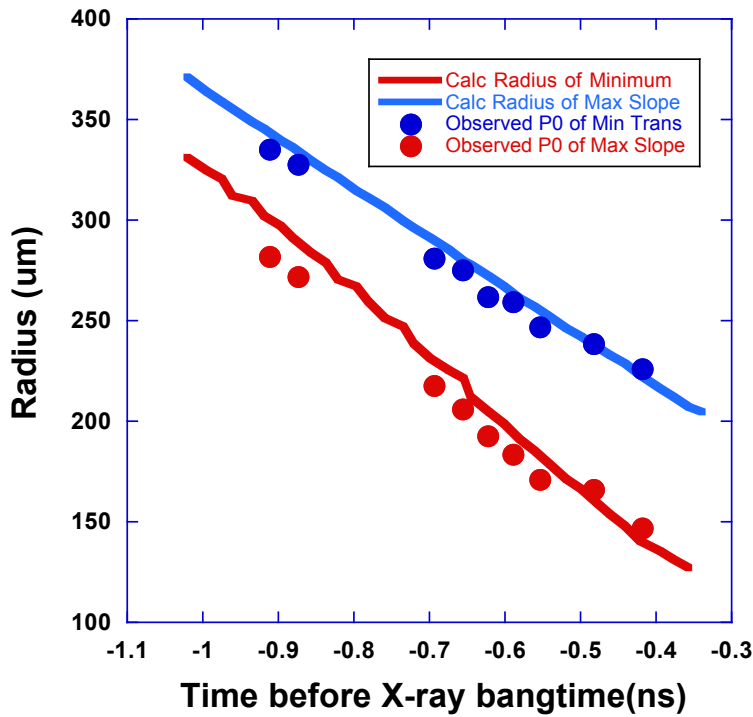

**Figure 7. Calculated and observed radii for the 2DConA implosion N150420**

Table I shows a comparison of post-shot calculations with observations for N150222 (1DconA), which forms the basis for modeling the DT layered shot. Both drive multiplier profiles, the two-step and multi-step can match the observed bangtime, x-ray self-emission width, DD ion temperature peak implosion velocity, mass remaining, and x-ray image radius, but the higher late time drive multipliers of the multi-step profile lead to a higher calculated yield.

**Table I- PostShot Simulations of N150222 vs Observations**

| Multiplier Type | Bangtime (ns) | Yield (e+11) | Xray fwhm (ps) | DD Tion (keV) | ConA peak V (km/s) | P0 (um) | P2/P0 (%) | Mass Remaining (%) |
|---|---|---|---|---|---|---|---|---|
| Two-Step (0.86) | 17.401 | 2.31 | 299 | 2.27 | 274 | 66.0 | 62.3 | 11 |
| Multi-Step | 17.437 | 6.28 | 300 | 2.17 | 285 | 67.9 | -10.3 | 11 |
| Data | 17.44±0.02 | 2.31±0.16 | 377±20 | 2.11±0.2 | 282±9 | ~60 | -16 | 10±1.5 |

Dante [18], an 18 channel soft x-ray spectrometer, uses filters, mirrors, and X-ray diodes to measure the absolute flux emitted out the laser entrance hole of the hohlraum between 50 eV and 20 keV. Because drive multipliers are applied to the laser power incident on the hohlraum, the flux and spectrum of the radiation field inside the hohlraum are affected. Dante measures this radiation emitted out the laser entrance hole (LEH) and is therefore sensitive to these multipliers. Using the two-step or the multi-step with its high late time multipliers makes s substantial difference, and is the reason the mult-step multipliers were chosen to rise toward the end of the pulse. Figures 8-10 compare calculated and observed Dante spectrally integrated fluxes for the 1DconA, the 2DconA, and the DT layered target. Also compared are the M-band fluxes, more specifically the flux integrated above 1.8 keV. There is excellent agreement except the M-band flux for the 2DconA. This is probably due to the lack of one beam in the Dante field of view which was diverted to the copper backlighter foil and to the presence of a large diagnostic hole where it would have hit the hohlraum wall.

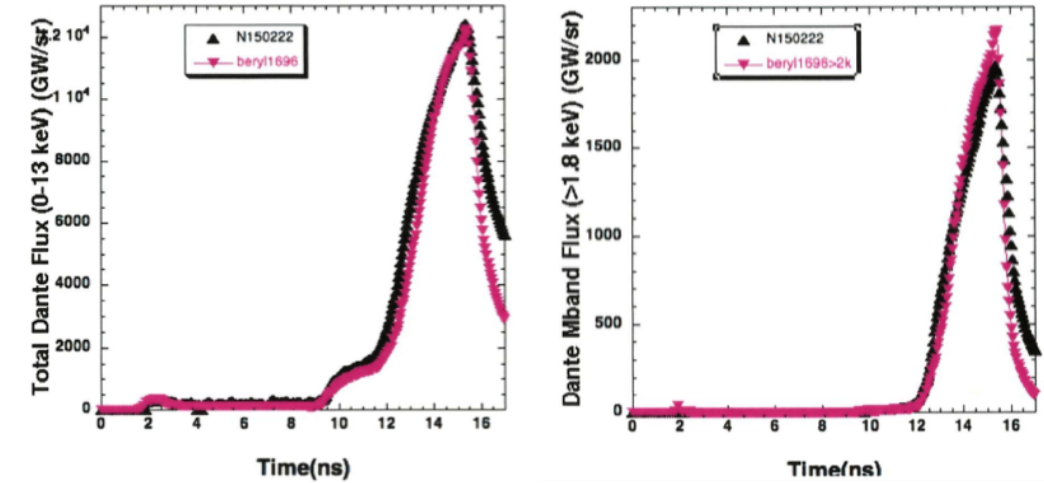

**Figure 8 Comparison of Dante measurements (black) with calculations (red) of total X-ray and high energy (>1.8keV or M-band) flux from shot N150222, a Be 1DconA target.**

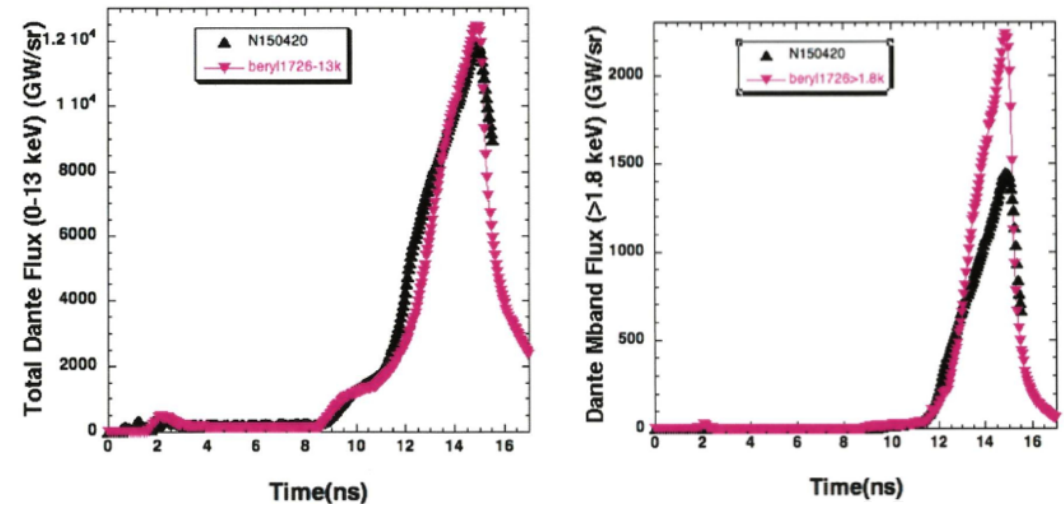

**Figure 9 Comparison of Dante measurements (black) with calculations (red) of total X-ray and high energy (>1.8keV or M-band) flux from shot N150420, a Be 2DconA target.**

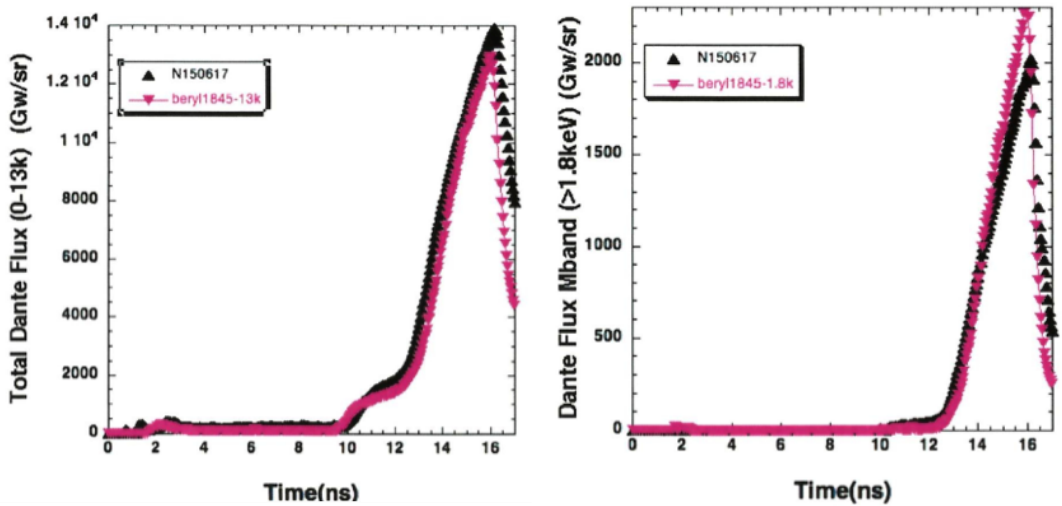

**Figure 10 Comparison of Dante measurements (black) with calculations (red) of total X-ray and high energy (>1.8keV or M-band) flux from shot N150617, a DT layered Be capsule.**

Since drive multipliers are thought to be a consequence of inadequate hohlraum, not capsule, modeling, the multi-step drive multipliers should apply to hohlraums with plastic (CH) as well as with beryllium capsules. Figures 11 and 12 compare calculated and measured Dante fluxes for two plastic DT layered capsule implosions, N130501 (1.26MJ) and N130812 (1.69MJ). Again black is the measured flux, but green is calculated with the multi-step multiplier, and red with the two-step. The high multipliers of the multi-step profile, cause an increase in the calculated fluxes, both total and m-band, bringing them into better agreement with the observations. For both of these, the 1$^{st}$ pulse and the second pulse, and the multiplier at the start of the 3$^{rd}$ pulse were taken from standard CH capsule models. Only the multipliers after 13ns were changed. The late time rise in the drive multiplier is essential for matching peak Dante fluxes.

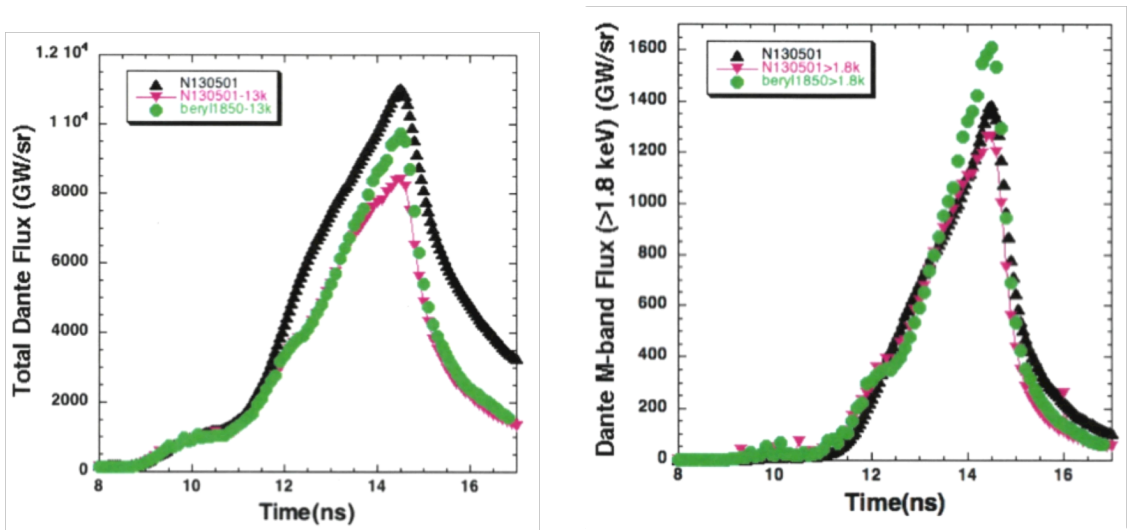

**Figure 11 Comparison of Dante measurements (black) with multi-step (green) and two-step calculations (red) of total X-ray and high energy (>1.8keV or M-band) flux from shot N130501, a DT layered CH capsule.**

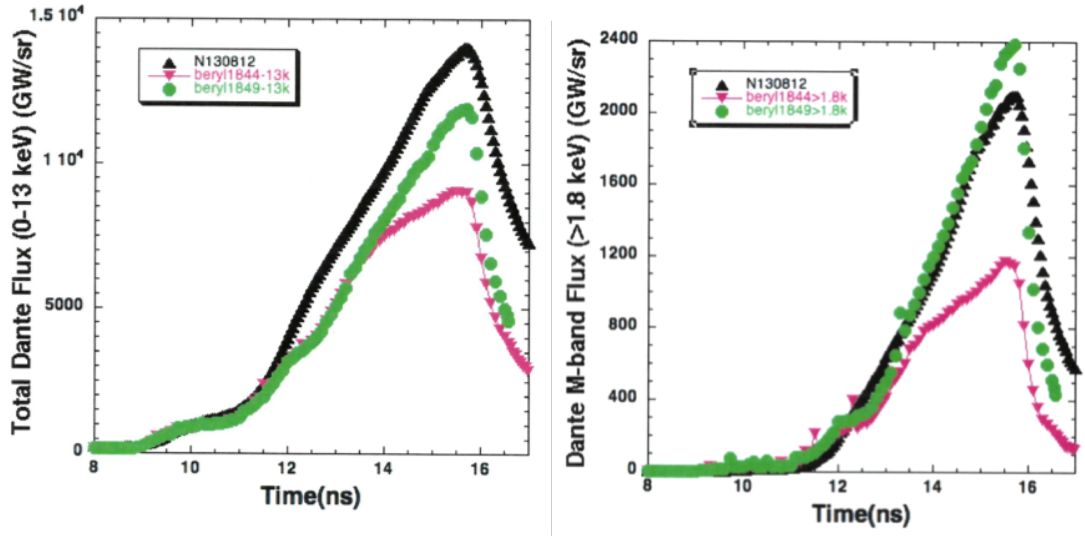

**Figure 12 Comparison of Dante measurements (black) with multi-step (green) and two-step calculations (red) of total X-ray and high energy (>1.8keV or M-band) flux from shot N130812, a higher laser energy DT layered CH capsule.**

The Dante instrument is not spatially resolved, but integrates all the X-ray emission coming within its field of view. Typically this emission comes from within the LEH, but the effective aperture of the LEH to X-rays is changing with time. Images of the LEH with the Static X-ray Imager (SXI) [19] can measure the amount of this closure, and quantify how well HYDRA calculations replicate it. Figure 13 shows the time integrated SXI images at 870eV (near the peak of a 300eV black body spectrum), both calculated and observed for the DT layered shot N150617. Both include a 2950 μm diameter circle representing the measured effective LEH diameter. The calculated image is smaller, and could be enclosed by a 2750 μm diameter circle. Thus the hohlraum brightness (flux/unit area) is about 13% less than calculated from the spatially integrated flux, or in a time integrated sense the multi-step laser drive multipliers are ~13% too high. Since the calculated and observed bangtimes agree, then a better drive multiplier set would rise earlier in the third pulse, peaking to a value to 1 at the end. This might also give better agreement with the 1DconA and 2DconA results, but this work has not yet been completed.

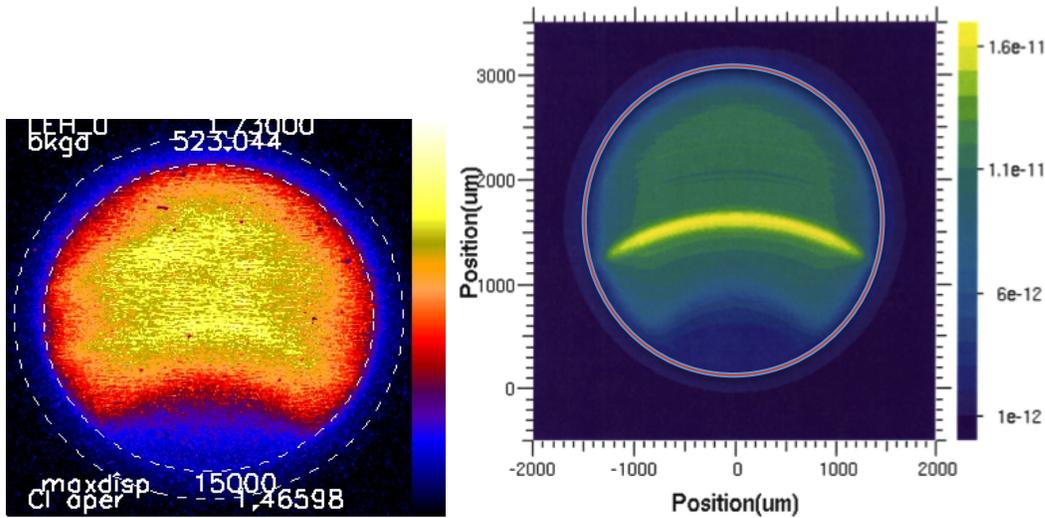

**Figure 13 Comparison of observed (left) and calculated (right) SXI images at 870 eV. Both the inner dotted circle on observed image and the pink circle on the calculated are 2950 μm diameter.**

## M-band flux in the laser picket is over-predicted by ~100 times

One of the deficiencies in the standard HYDRA modeling of ICF capsules is the over-prediction of M-band flux in the picket of the laser pulse. Drive multipliers cannot correct for this. The drive multiplier in the picket is set to obtain the 1$^{st}$ shock velocity observed by the VISAR, 0.93 for beryllium capsules and 0.85 for CH. This multiplier is thought to represent errors in the equation of state for each material. With the multiplier fixed, Figure 14 shows a comparison of calculated and observed Dante M-band fluxes. The M-band is over-predicted for both capsule types, suggesting the error is in hohlraum, not capsule modeling. HYDRA calculates that a thin hohlraum wall layer is being heated to high temperature at low density. The data suggest rather that softer X-rays are being emitted from lower temperature plasmas. The total Dante flux seems correctly calculated. Other than artificially inhibiting the expansion of the ablator/fuel interface, we know of no way of correcting for this error in an integrated calculation.

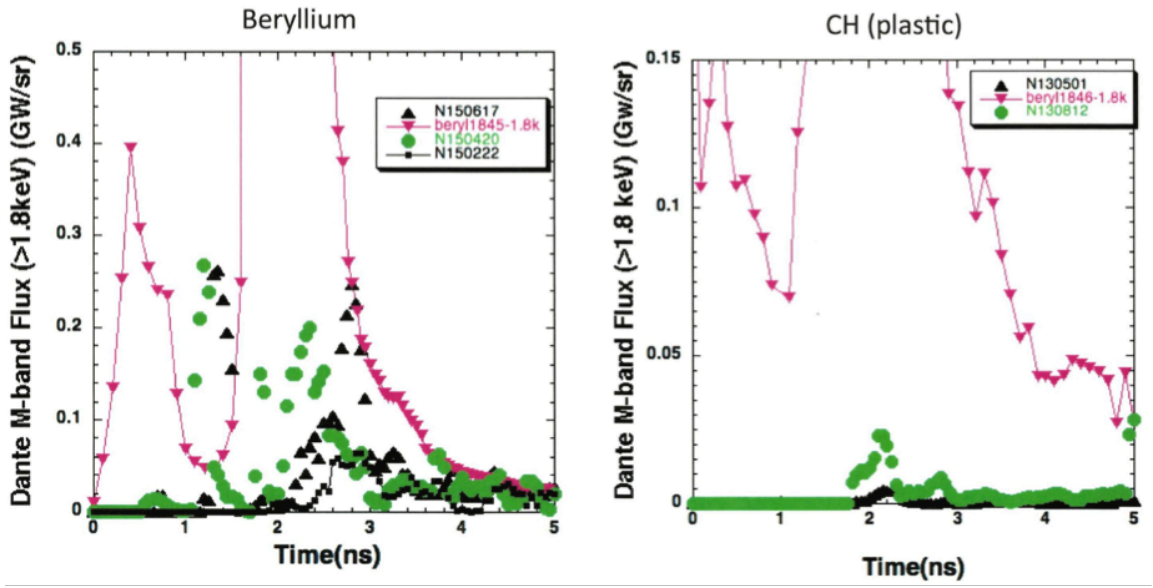

**Figure 14 Comparison of Dante measurements (black and green) with calculations (red) of high energy (>1.8keV or M-band) flux from hohlraums with Be capsules (left) and CH capsules(right).**

For beryllium capsules this over-prediction has a measureable consequence in VISAR measurements. Unlike CH which transmits VISAR light, the metallic beryllium/DD surface reflects the VISAR. M-band X-rays can penetrate through the ablator, preferentially heating the copper doped layers. These then expand to drive a shock to cause a measureable inward motion of the beryllium/DD interface. HYDRA calculations predict an interface velocity of 2-4 km/s. VISAR measurements on shot N140829 show an expansion of only 0.2-0.3 km/s, reflecting about 1/100 the kinetic energy, consistent with ~ 1% of the predicted M-band flux.

## Laser Power Delivery Errors in the Picket Affect the Implosion

With the multi-step drive multipliers determined from VISAR, 1DconA and 2DconA experiments, the hohlraum radiation drive could be calculated for the DT layered beryllium shot N150617. However a small but important difference occurred between the intended laser picket and that delivered by the NIF laser. The delivered energy in the picket was 9% low (within NIF specifications), but the second and third pulses were as requested. This caused a slower first shock in the beryllium. The second and third shocks then over took the first shock 20 μm inside the DTice/ gas interface, instead of just inside the gas. This increased the entropy of the fuel, changing the total fuel adiabat from an intended 1.9 to 2.3 times the Fermi degenerate. The final ρR was decreased. The expected down-scattered ratio (DSR) dropped from 5% to 4%, and the calculated yield was reduced a factor of 2. This unfortunate picket deficit affected the first CH DT layered capsule as well, N130501, allowing us to still compare the relative performance of beryllium to plastic.

## Implosion Symmetry

The symmetry of an implosion is governed by the placement of laser deposition on the hohlraum wall. The pointing of the laser, the fraction of the energy in the inner versus outer laser beams (cone fraction), and motion of the critical surface away from the wall and capsule, all affect the symmetry. The pointing and time dependent cone fractions are tuned during the campaign of keyhole, 1DconA, and 2DconA targets. One of the most important factors affecting symmetry is the transfer of energy between beams at different incident angles as they pass through the laser entrance hole via induced Brillouin scattering [20]. Experimental tuning is accomplished by changing the relative laser wavelengths of the beams by a few angstroms. Symmetry along the hohlraum axis, as measured by self emission and transmitted X-rays, was corrected by adjusting the wavelength of the 23° cone beams to be 5.9 angstroms more than the outer cone beams (44°and 50°). The staggering in azimuth of the inner cone beams allowed a correction for azimuthal asymmetry such that the 30° beams were at a 0.7 angstrom shorter wavelength. The post-processing modeling of cross-beam transfer (CBET) in HYDRA is approximate, and requires limiting the beam transfer in the third laser pulse by a saturation of the acoustic wave amplitude to successfully model observed X-ray symmetry. While a similar saturation model is used for all the beryllium capsule implosions, the high sensitivity to the saturation parameter meant that modeling a specific implosion was *ad hoc*. With a model saturation parameter of 2.5 e-4., the azimuthal asymmetry of DT layered implosion is well approximated, as shown below.

## DT Implosion Results

The DT layered capsule culminating the series of NIF beryllium experiments was shot on June 17, 2015. The laser delivered 1420 kJ while the total SRS and SBS backscatter was 234.8 kJ so that 83.26% of the laser energy was coupled to the target. The capsule produced $7.77 \pm 0.124$ e+14 13-15 MeV neutrons, a DT ion temperature of $3.65 \pm 0.13$ keV, a DD ion temperature $3.53 \pm 0.17$ keV, and a down scattered ratio of $3.2 \pm 0.23$ %. The xray bangtime (time of peak xray emission) was 17.622ns, the xray emission width fwhm $130 \pm 13$ps, while the gamma ray burn width was $142 \pm 30$ps fwhm. The 2D post-shot simulation gave 2.67e+15 13-15 MeV neutrons, a DT ion temperature of 3.27 keV, and a down-scattered ratio of 3.95%. Thus the measured yield was 29% of simulated with alpha deposition. Without any energy from alpha deposition the 2D yield calculates to be only, 1.77e+15. If the implosion had been perfectly symmetric the calculated yield without alpha deposition would have been 2.70e+15 (13 to 15 Mev neutrons) with a DT ion temperature of 3.04keV and with alpha deposition, 5e+15 and 4.5 keV.

### X-ray and Neutron Imaging

Time integrated images taken in hard X-rays [21] reflect the emission of hot, translucent or opaque material within the capsule. If the X-rays are sufficiently high energy, greater than ~ 7 keV, then they are weakly attenuated and reflect the symmetry of the hot DT fuel and surrounding material. Figure 14 and Table II compare the measured and calculated X-ray emission images from detectors on the equator of the hohlraum axis. Figure 15 and Table III compare them from the north polar direction. Since the calculations are 2D we cannot capture azimuthal variation around the polar axis, or left-right asymmetry in the equatorial images.

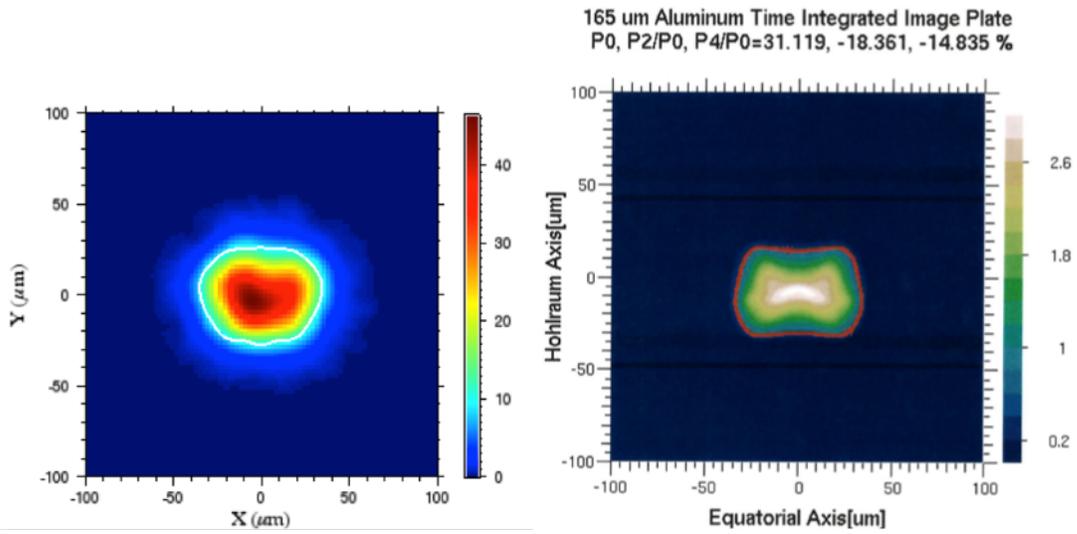

**Figure 14** Measured (left) and calculated (right) X-ray image plates from the φ=78° equatorial direction. Hohlraum axis is vertical.

**Table II. Equatorial X-ray Image Shapes**

|  | Measured | Calculated Post Shot |
|---|---|---|
| **PSL (signal level)** | 14868 | 5407 |
| **P0 (µm)** | 31.9 | 31.1 |
| **P2/P0 %** | -15.89 | -18.4 |
| **P4/P0 %** | -2.69 | -14.8 |

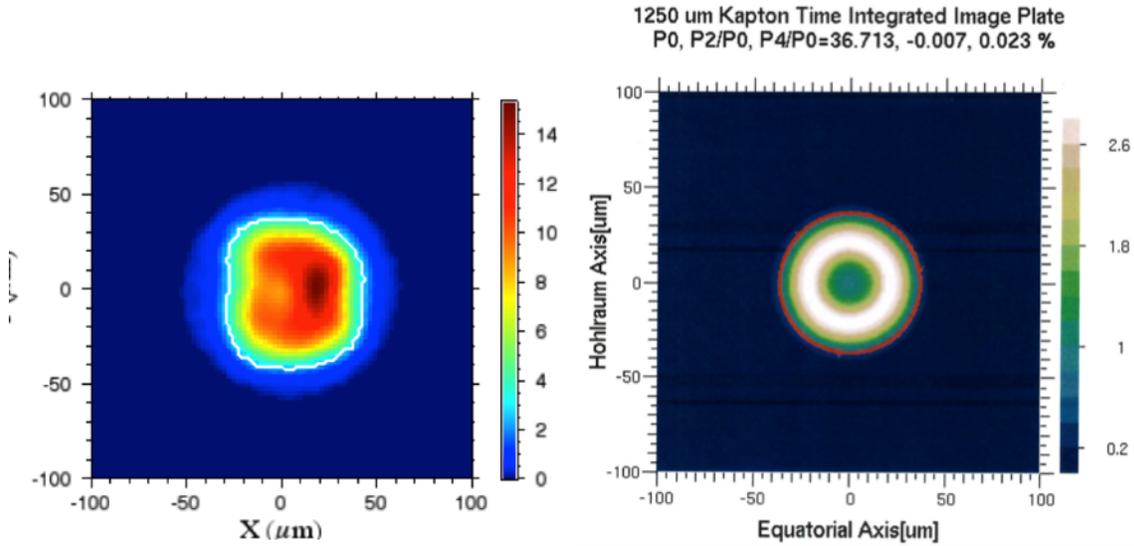

**Figure 15** Measured (left) and calculated (right) X-ray image plates from the north polar direction.

Table III. Polar X-ray Image Shapes

|  | Measured | Calculated Post Shot |
|---|---|---|
| **PSL (Signal)** | 7368 | 9276 |
| **M0($\mu$m)** | 37.1 | 36.7 |
| **M2/M0 %** | 5.34 | 0 |
| **M3/M0 %** | 3.02 | 0 |
| **M4/M0%** | 5.47 | 0 |
| **M2 Phase** | 96° | 0 |
| **M4 Phase** | 43° | 0 |

Images in 13-15 MeV neutrons [22], taken on the equator at 315° in azimuth, reflect symmetry of the burning DT unaffected by attenuation or scattering. The radius of the contour of 17% of peak brightness is analyzed in Legendre moments for both X-ray and neutron images. The radius, or P0, is 32 μm in both the equatorial X-ray and neutron images and both show a oblate character (P2/P0 =-16 and -19%). The agreement in P0 with the calculation arises from using the correct hohlraum radiation drive, which has been adjusted with laser drive multipliers. The agreement in shape arises from the choice of saturation of beam transfer. Neutrons from 6-12 MeV have been scattered once from their initial energy near 14 MeV. The images taken using these neutrons reflect the location of those scatterers, primarily the DT fuel itself. Thus the calculated and observed images in figure 17 show the location and asymmetries of all the fuel, which is dominated by the relatively cold fuel in the shell surrounding the hot spot. This is evident in figure 18 which overlays the hot spot (13-15 MeV) image onto the down-scattered image. Consequently the radius (P0) of the down-scattered image is larger than the direct 13-15 MeV image. Also note the triangular shape of the down-scattered image, indicating significant 3D asymmetry.

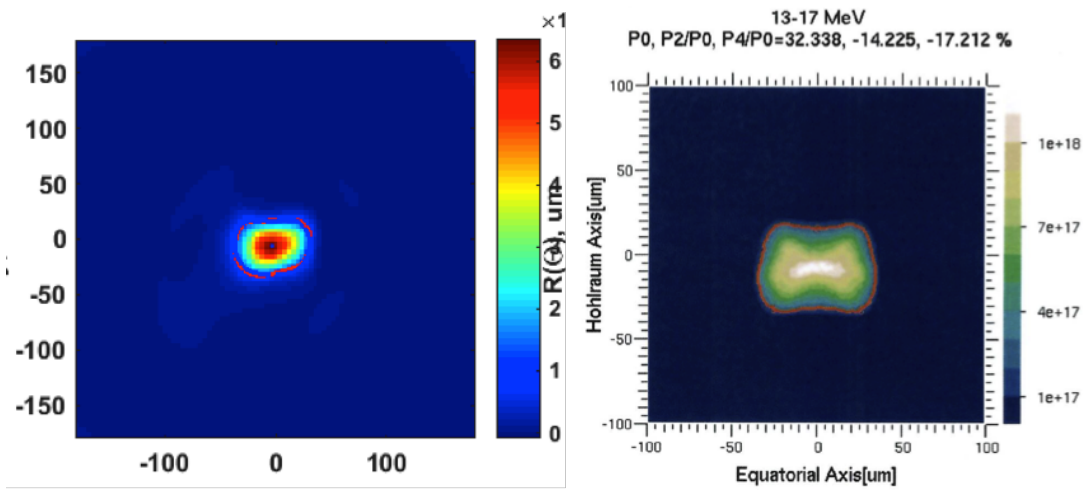

**Figure 16 Measured equatorial 13-17 MeV neutron image φ=315° (left) and simulated image at approximately the same scale.**

**Table III. Equatorial 13-17 MeV Neutron Image Shapes**

|  | Measured | Calculated Post Shot |
|---|---|---|
| **P0 (μm)** | 32 | 32.3 |
| **P2/P0 %** | -19 | -14.2 |
| **P4/P0 %** | -7 | -17.2 |

Figure

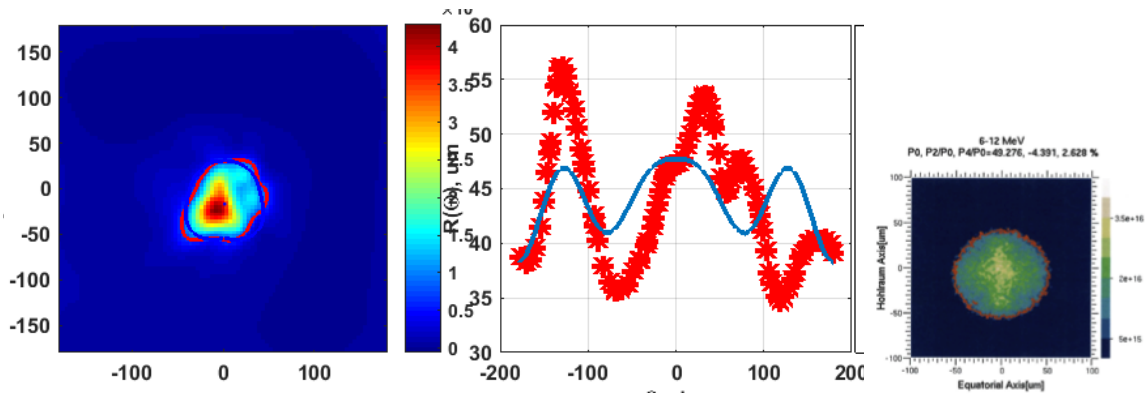

**Figure 17 Measured equatorial 6-12 MeV neutron image at φ=315° (left) and simulated image at approximately the same scale. The blue line represents the Legendre analyzed contour, poorly representing the actual (red) contour.**

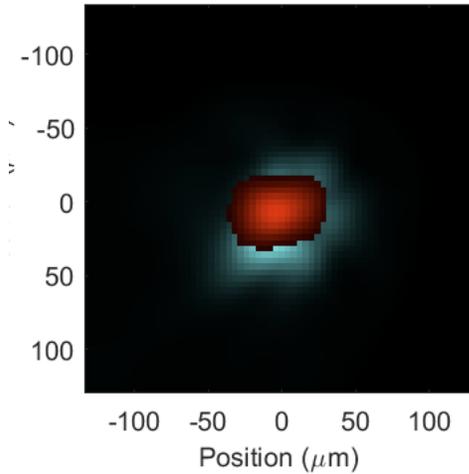

**Figure 18** Overlay of the observed down-scattered 6-12 MeV neutron image (blue) with primary the 13-17 Mev image (red).

**Table IV. Equatorial 6-12 MeV Neutron Image Shapes**

|  | **Measured** | **Calculated Post Shot** |
|---|---|---|
| **P0 (μm)** | 44 | 49 |
| **P2/P0 %** | 5 | -4.4 |
| **P4/P0 %** | -7 | 2.6 |

## Measurements of ρR

The areal density of DT within a compressed capsule can be measured using 14 MeV DT neutrons that have scattered once in the target. These are collected between 10 and 12 MeV. The ratio of these to the un-scattered neutrons, between 13 and 15 MeV, is approximately proportional to the areal density of the DT fuel, the dominant scatterer. This ratio, the DSR, is calculated to be 3.78 %, comparing well with 3.8 ± 0.03% measured by the Magnetic Recoil Spectrometer (MRS). The neutron time of flight (NTOF) measurements from several instruments gave 2.92 ± 0.19 %. Often this is hypothesized to be due being measured at different polar and azimuthal angles in the target chamber. The top/down asymmetry in laser energy does give a polar angle dependence , but this cannot account for the difference between the MRS and NTOF measurements, which we currently do not understand. Figure 19 shows the Flange Nuclear Activation Detector (FNAD) analysis[23]. Because of the decreasing neutron activation cross section of Zirconium below ~12.5 MeV, the ratio of the activation yields at different locations in the target chamber reflects the differing amounts of scattered neutrons in each line of sight. Figure 19 graphically shows a fit to the angular distribution of relative yields from 18 FNAD detectors. Higher areal density is present at the north pole than the south. There is also lower areal density near Φ=45°, θ=90°. This may be due to an initial M1 asymmetry in the DT ice layer at formation, measured with a 2.1 μm peak to valley thickness variation with the thick side near Φ=200°. This 3D asymmetry is not calculated. Figure 20 compares the relative FNAD yield with polar angle to the calculated DSR values.

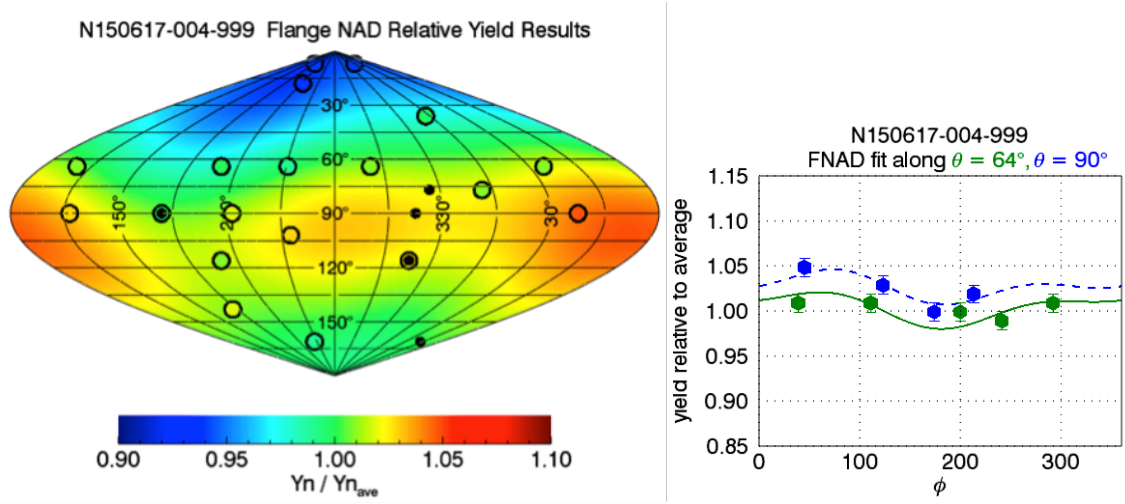

**Figure 19 FNADs neutron activation around the target chamber.**

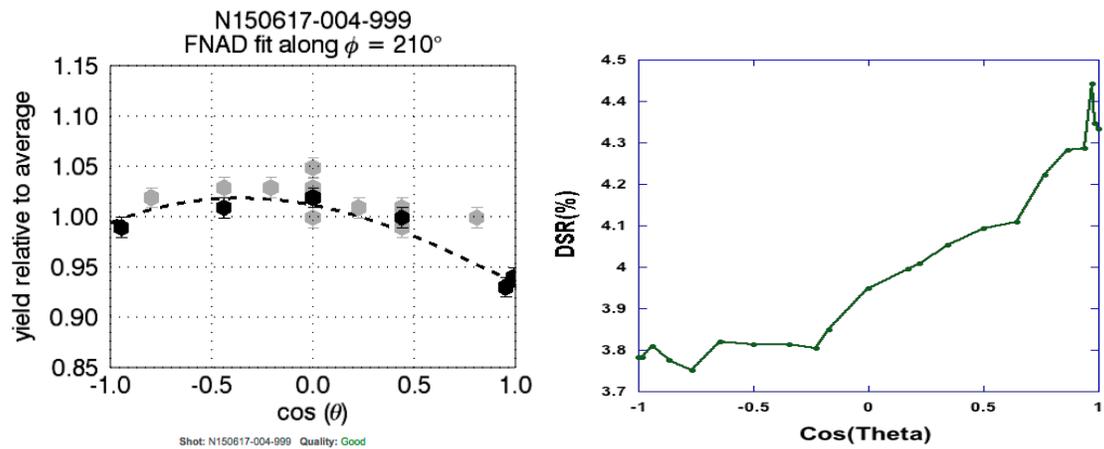

**Figure 20 FNADs neutron activation variation with polar angle at Φ=210° compared to calculated DSR variation.**

## $^{41}$Ar Activation

Because rare gas sampling had been included, the implosion yielded a serendipitous measurement of the first activation of argon at NIF. The $^{41}$Ar/$^{40}$Ar(capsule)/13-15 MeV neutron ratio is a measure of beryllium ρR during DT burn. 1.15e-7 is calculated, representing a Be ρR of 0.360 g/cm$^2$ during burn. Preliminary analysis shows that we obtained 11.4 ± 3.2 e+7 $^{41}$Ar atoms, or 1.46e-7 atoms/neutron, in agreement with the calculation. Suggesting again that our calculation of the hohlraum energetics is correct and that there is no un-calculated preheat to the beryllium which would decrease its areal density.

## Degradation by High Mode Perturbations

Any perturbations to the beryllium surface could lead to a degradation of DT neutron yield. Simulations show mid-mode roughness can give an additional 10-30% degradation in post-shot yield (assuming 1-3x specified roughness). Figure 21 shows how the density and ion temperature are perturbed in 2D-capsule only calculations taking into account P2 and P4 drive asymmetry, and the measured beryllium and DT ice roughness up to mode 100. The right figure shows how mid-mode roughness degrades the yield in addition to the drive degradation.

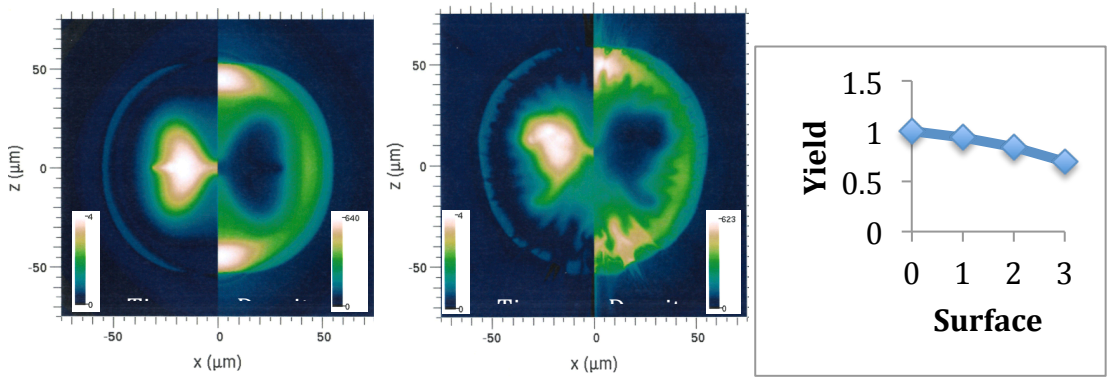

**Figure 21 The effects of calculated surface roughness on ion temperature and density profiles at peak yield rate (no roughness (left), 3X measured (center)), and the effect on yield of surface roughness multipliers (right).**

# Conclusions

Table V compares measured and inferred quantities for one Be and three CH NIF DT shots. Yields are calculated either with or without an enforced radiation drive symmetry, and with or without alpha deposition. In the absence of heating from alpha deposition a symmetric radiation drive accounts for up to a factor of two degradation. This degradation leads to less alpha particle deposition and less heating. It accounts for N130812 not igniting. From the laser energy, measured DSR, implosion velocity, remaining mass (in % of initial or in μg), and ratio of observed yield/ calculated 2D asymmetric, we see that beryllium and plastic implosions are quite similar. The laser-hohlraum interaction is only slightly different with Be than CH capsules. Drive multipliers for the 3$^{rd}$ pulse are similar, maybe, identical for Be and CH. X-ray preheat in the 1$^{st}$ pulse is wildly over-calculated (by ~ 100X) for both Be and CH hohlraums. Similar saturation of cross-beam energy transfer is needed for both Be and CH capsule hohlraums to replicate observed x-ray and neutron images. In summary when driven similarly the Be capsules performed like CH capsules. Performance degradation for both seems dominated by drive and capsule asymmetries.

**Table V. Be and CH implosion drives and inferred Results**

|  | N150617 | N130501 | N130710 | N130812 |
|---|---|---|---|---|
|  | Be | CH | CH | CH |
| **Laser Energy (MJ)** | 1.403 | 1.268 | 1.50 | 1.687 |
| **DSR (MRS)** | 3.8% | 3.07% | 3.4% | 4.0% |
| **Fuel Velocity (km/s)** | $342^2$ | $308^2$ | $340^2$ | $353^2$ |
| **Remaining Mass (µg)** | 200(5%) | 420(15%) | 310(11%) | 280(10%) |
| **Yield ($10^{14}$)** | 7.87 (29%)[1] | 7.71 (29%)[1] | 10.5(21%)[1] | 23.8(21%)[1] |
| Asym, no α | 15.0 | 8.91 | 25.5 | 39.9 |
| Asym, with α | 26.8 | 11.2 | 50.2 | 112 |
| Symm, no α | 26 | 16.8 | 44.6 | 70.3 |
| Symm, with α | 54.6 | 17.1 | 147 | 5550 |
| Asym/Symm | **49%** | **65%** | **34%** | **2%** |

[1] Observed / Unmixed 2D Asymmetric Drive with Alpha Deposition

[2] Hurricane *et al*. [9] inferred 297, 323 and 307 ± 15 km/s using lower drive multipliers

## Acknowledgements

Execution of NIF experiments requires the participation and collaboration of large teams of individuals.  We would like to thank all of those who make the NIF laser a success, in particular the experimentalists who created and fielded the diagnostics, the target fabrication specialists who performed the myriad tasks involved in designing, building, and fielding the complex holraums and beryllium capsules in our experiments.  This work was performed at Los Alamos and Lawrence Livermore National Laboratories, funded by the US Department of Energy.